\documentclass{aa}
\usepackage{graphicx}
\begin{document}

%
   \title{The IR counterpart of the black-hole candidate
   4U\,1630$-$47\thanks{Based on observations collected at the European
   Southern Observatory, Chile.}}

   \author{T. Augusteijn\inst{1}
          \and
          E. Kuulkers\inst{2,3}
          \and
          M. H. van Kerkwijk\inst{3}
          }

   \offprints{T. Augusteijn, \email{tau@ing.iac.es}}

   \institute{Isaac Newton Group of Telescopes,
              Apartado de Correos 321, 38700 Santa Cruz de La Palma, 
              Canary Islands, Spain 
         \and
              Space Research Organization Netherlands, Sorbonnelaan 2, 
              3584 CA Utrecht, The Netherlands
         \and
              Astronomical Institute, Utrecht University, P.O. Box 80000, 
              3508 TA Utrecht, The Netherlands
             }

   \date{Received 5 June 2001 / accepted 19 June 2001}

   \titlerunning{The IR counterpart of 4U\,1630$-$47}
   \authorrunning{T.~Augusteijn et al.}

\abstract{We present $K$ band photometry of the region including the radio
error box of the soft X-ray transient 4U\,1630$-$47 during its 1998
outburst. We detect a variable source at $K$=16.1 mag located inside the
radio error circle which we identify as the counterpart to the X-ray
source. We discuss the properties of the source, and conclude that it is
most likely a black-hole X-ray binary similar to \object{4U\,1543$-$47},
\object{GRO\,J1655$-$40} or \object{SAX\,J1819.3$-$2525}, containing a
relatively early-type secondary.
\keywords{Accretion, accretion disks -- {\bf Stars: individual:}
4U\,1630$-$47 -- X-rays: stars} }

   \maketitle

%

\section{Introduction}

Despite the regular outburst behavior (600--690~days, e.g., Jones et al.
\cite{jones76}; Kuulkers et al. \cite{gnotor97b}) of the soft X-ray
transient (SXT) \object{4U\,1630$-$47}, little is known about this source.
Its X-ray spectral (Parmar et al. \cite{parmar86}; Barret et al.
\cite{barret96}; Oosterbroek et al. \cite{tim98}) and X-ray timing (Kuulkers
et al. \cite{gnotor97a}; Sunyaev \& Revnivtsev \cite{sunyaev00}) properties
during outburst suggest it harbors a black-hole (see also Dieters et
al. \cite{stefan00}; Tomsick \& Kaaret \cite{tomsick00}; Trudolyobov et
al. \cite{trudo01}).

Optical/infra-red identification of 4U\,1630$-$47 has been hampered due to
the high extinction and the crowded field near its X-ray position (Reid et
al.  \cite{reid80}; Parmar et al. \cite{parmar86}). A more precise location
became available when the source was discovered in the radio during its 1998
outburst (Hjellming \&\ Kuulkers \cite{hjell98}; Buxton et al.
\cite{buxton98}; see also Hjellming et al. \cite{hjell99}): (J2000) RA =
$16^h34^m01\fs6 \pm 0\fs05$, Dec = $-47\degr23\arcmin33\arcsec \pm 2\arcsec$
(VLA) or RA = $16^h34^m01\fs61 \pm 0\fs02$, Dec =
$-47\degr23\arcmin34\farcs8 \pm 0\farcs3$ (ATCA). This position lies well
within the Einstein HRI error region (Parmar et al. \cite{parmar97}).

A first attempt to localize the infrared counterpart during the 1996
outburst was recently presented by Callanan et al. (\cite{callan00},
henceforth CMG). We here report on a search for the infrared counterpart
during its 1998 outburst.

\section{Observations and reduction}

We observed 4U\,1630$-$47 with the infrared spectrograph and imaging camera
SOFI on the 3.5-m NTT at the European Southern Observatory, La Silla, Chile
on 4 separate occasions during the course of the 1998 outburst. A log of the
observations is presented in Table~\ref{tab:log}.

All observations were made with a Rockwell HgCdTe 1024$\times$1024 Hawaii
array using the ``small field'' objective, which provides a 2\farcm
47$\times$2\farcm 47 field, with a pixel size of 0\farcs 144. The $K_s$
filter used during the observations is slightly different from both the $K$
and $K^{\prime}$ filters defined by Wainscoat \& Cowie (\cite{wain92}). The
long wavelength edge of the $K_s$ is similar to that of the $K^{\prime}$
filter, but the short wavelength edge is similar to that of the $K$
filter. Thus, the $K_s$ filter avoids both the atmospheric absorption
feature at 1.9 ${\mu}$m and the radiation from the thermal background beyond
2.3 ${\mu}$m (Lidman \& Cuby \cite{lidman98}).

To accurately determine the sky brightness across the field the observations
in each night consisted of 20 slightly-dithered frames with an exposure time
of 30 seconds each. These frames were flat-fielded and sky subtracted. The
source is located in the Galactic plane, and crowding is the main limitation
to reaching faint magnitudes. We therefore selected in each case the best
quality images and co-registered these to form a master frame for each
night, producing a total on-target integration time of between 270 and 330
seconds. The seeing in the resulting frames as derived from the
point-spread-function (PSF) determined for each observation (see
Sect.~\ref{subsect:phot}) is listed in Table~\ref{tab:log}.

\begin{table}[t]   
\begin{center}
\caption{Observing log}
\label{tab:log}
\begin{tabular}{cccc} 
\hline
\multicolumn{1}{c}{Date (UT)} & \multicolumn{1}{c}{T$_{\rm mid}$(JD$-$2450000)} & \multicolumn{1}{c}{T$_{\rm exp}$(sec)} & \multicolumn{1}{c}{seeing(\arcsec)} \\ \hline
12/03/1998 & 885.90196 & 300 & 1.00 \\
14/03/1998 & 887.90360 & 270 & 0.59 \\
18/03/1998 & 889.90981 & 330 & 0.51 \\
16/05/1998 & 950.83178 & 300 & 0.75 \\ \hline
\end{tabular}
\end{center}
\end{table}

\begin{figure*}
 \resizebox{12cm}{!}{\includegraphics{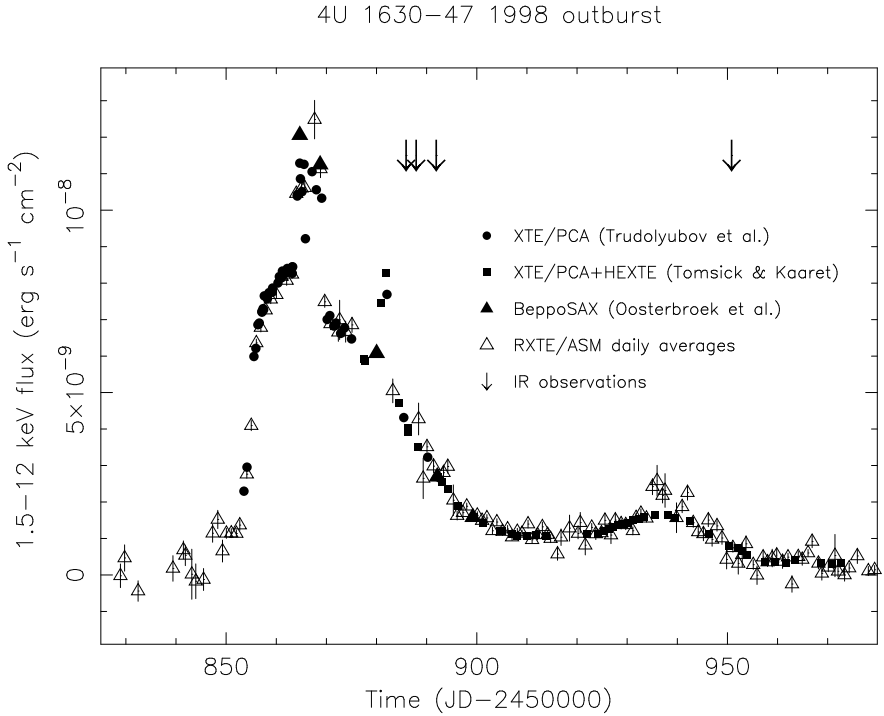}}
 \hfill
 \parbox[b]{55mm}{ 
 \caption{ The observed 1.5--12\,keV fluxes of 4U\,1630$-$47 during the 1998
 outburst. The arrows indicate the mid-times of the IR observations } 
 \label{fig:xrlc}}
\end{figure*}

\section{X-ray outburst light curve}

In Fig.~\ref{fig:xrlc} we give the observed 1.5--12\,keV fluxes (i.e.\ not
corrected for absorption) during the 1998 outburst of 4U\,1630$-$47 (January
11 to June 15, 1998). The RXTE/ASM data points were derived by dividing the
observed daily averaged count rates by the total average Crab count rate
during the same period (74.94$\pm$0.20\,ASM cts\,s$^{-1}$) and assuming that
1~Crab equals 2.975$\times$10$^{-8}$\,erg\,s$^{-1}$\,cm$^{-2}$
(1.5--12\,keV). The other data points were derived from the spectral fit
parameters as given by Oosterbroek et al.\ (\cite{tim98}: BeppoSAX/MECS+PDS,
their Table~2), Tomsick \&\ Kaaret (\cite{tomsick00}: RXTE/PCA+HEXTE, their
Table~5) and Trudolyubov et al.\ (\cite{trudo01}: RXTE/PCA, their
Table~2). We have normalized the derived fluxes from the observations of
Oosterbroek et al.\ (\cite{tim98}) and Trudolyubov et al.\ (\cite{trudo01})
to those of Tomsick \&\ Kaaret (\cite{tomsick00}) using observations done
closest to each other in time (i.e.\ near JD\,2450892 or March 19,
1998). This resulted in correction factors of 1.30 and 0.91 for the
Oosterbroek et al.\ (\cite{tim98}) and Trudolyubov et al.\ (\cite{trudo01})
observations, respectively. In Fig.~\ref{fig:xrlc} we have also indicated
the mid-times of the IR observations discussed below with arrows.

The outbursts in 4U\,1630$-$47 occur fairly regularly. From its first
detection in 1969 to the mid-1980's the outbursts appeared to be periodic
with a recurrence time of $\sim600$~days. However, from the mid-1980's to
the 1998 outburst discussed here the recurrence time appeared to be longer,
at about 690~days (Kuulkers et al. \cite{gnotor97b}; Kuulkers
\cite{gnotor98b}). The latest two outbursts, which reached maximum X-ray
fluxes as derived from the RXTE/ASM data at approximately JD~245\,1350 and
245\,1885, respectively, indicate a different recurrence time again. A
linear fit to all dates indicates an average recurrence time of
$\sim610$~days, with a root-mean-square (rms) spread of $\sim100$~days. We
believe that the observed variation in the outburst timings is in fact
intrinsic to the mechanism causing the outbursts, and the recurrence time
just reflects the typical time-scale. The next outburst is expected to occur
somewhere in the second half of the year 2002.

\section{Search for a Near-Infrared Counterpart}

\subsection{Astrometry}

Astrometry was done relative to the USNO-A2.0 catalogue (Monet et al.\
\cite{monet98}). For all 35 USNO-A2.0 stars overlapping with the best-seeing
image of March 18, 1998, centroids were determined. Five of these were
extended and therefore rejected. With the remaining stars, the zero point
position, the plate scale, and the position angle on the sky were
determined. After rejection of four outliers, with positions deviant by more
than 0\farcs5, the rms residuals for the 26 remaining stars were 0\farcs15
and 0\farcs19 in right ascension and declination, respectively, consistent
with the expected uncertainties on the USNO-A2.0 positions. Therefore, we
expect that the tie to the system defined by USNO-A2.0 is good to about
0\farcs04. The uncertainty in the tie to the International Celestial
Reference Frame should thus be dominated by the systematic uncertainties in
USNO-A2.0 itself, of about 0\farcs2.

\begin{figure*}
\resizebox{\hsize}{!}{\includegraphics{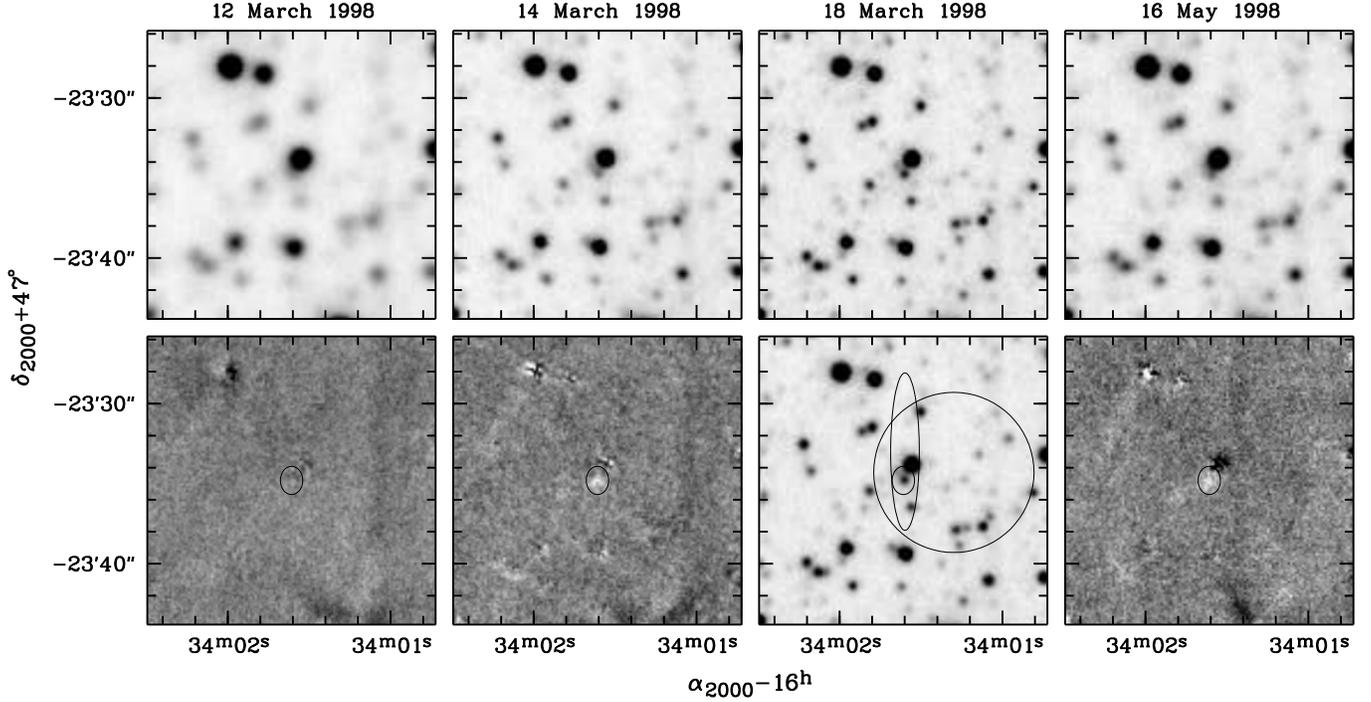}}
\caption[]{$K$ band images of the field of 4U\,1630$-$47, taken at the dates
indicated. Each image covers an area of $\sim$18\arcsec$\times$18\arcsec . For
the 18 March image, the lower panel shows the image with error circles from,
in order of increasing area, ATCA, VLA, and HRI. For the other dates, the
lower panels show difference images with the 18 March (best-seeing)
image. There is a variable source in the error circle, which is brighter at
12 and 18 March than on 14 March and 16 May.}
\label{fig:irvar}
\end{figure*}

The other images were rebinned to an image matching in position and
orientation with the 18 March, 1998 image, using some 40 secondary stars to
determine the alignment. The aligned images are shown in the top row in
Fig.~\ref{fig:irvar}. Also shown are the 95\% confidence error circles
derived using the ATCA, VLA and HRI positions (where for each position, the
uncertainty we used includes the 0\farcs2 uncertainty in the optical
position added in quadrature). Clearly, several stars lie inside the
smallest (ATCA) error circle.

\subsection{Photometry}\label{subsect:phot}

\begin{table*}[]   
\begin{center}
\caption{$K$ band photometry}
\label{tab:IRphot}
\begin{tabular}{rrrrlrlllrl} 
\hline
\multicolumn{1}{c}{No} & \multicolumn{1}{c}{$\frac{\Delta\alpha}{cos\;\delta}$} & \multicolumn{1}{c}{$\Delta\delta$} & \multicolumn{2}{c}{12/03/1998$^b$} & \multicolumn{2}{c}{14/03/1998$^b$} & \multicolumn{2}{c}{18/03/1998$^c$} & \multicolumn{2}{c}{16/05/1998$^b$}\\ 
& \multicolumn{2}{c}{(arcsec)$^a$} & \multicolumn{1}{c}{$\Delta K$} & \multicolumn{1}{c}{$\sigma_{\Delta K}$} & \multicolumn{1}{c}{$\Delta K$} & \multicolumn{1}{c}{$\sigma_{\Delta K}$} & \multicolumn{1}{c}{$K$} & \multicolumn{1}{c}{$\sigma_{K}$} & \multicolumn{1}{c}{$\Delta K$} & \multicolumn{1}{c}{$\sigma_{\Delta K}$}\\ \hline
 1 & --0.50 &   0.98 &   0.000 & 0.025 & --0.007 & 0.011 & 13.630 & 0.011 & --0.035 & 0.033  \\ 
 2 &   0.00 &   0.00 &   0.058 & 0.041 &   0.215 & 0.032 & 16.011 & 0.018 &   0.198 & 0.041  \\
 3 & --0.44 & --1.67 &   0.112 & 0.036 &   0.024 & 0.025 & 16.056 & 0.019 &   0.001 & 0.039  \\
 4 &   0.51 &   0.54 &   0.351 & 0.091 & --0.051 & 0.072 & 17.118 & 0.033 &   0.120 & 0.069  \\
 5 & --0.63 & --0.56 &   0.17  & 0.12  &   0.29  & 0.10  & 17.520 & 0.063 &   0.074 & 0.089  \\
 6 &   1.42 & --0.03 & --0.36  & 0.17  &   0.05  & 0.13  & 17.823 & 0.070 & --0.03  & 0.13   \\
 7 & --1.73 &   1.11 &    ..   &  ..   &   0.42  & 0.15  & 17.861 & 0.072 & --0.01  & 0.12   \\
 8 &   1.40 &   1.84 & --0.18  & 0.14  &   0.24  & 0.12  & 17.971 & 0.055 &   0.20  & 0.12   \\
 9 & --0.02 &   0.51 & --0.79  & 0.20  & --0.23  & 0.24  & 18.25  & 0.12  & --0.38  & 0.21   \\
10 & --1.72 & --1.92 &   0.22  & 0.20  &   1.28  & 0.42  & 18.36  & 0.11  &   0.55  & 0.23   \\
11 &   1.19 & --0.85 & --0.89  & 0.24  & --0.25  & 0.21  & 18.53  & 0.14  & --0.14  & 0.27   \\
12 &   0.56 & --1.09 &    ..   &  ..   &    ..   &  ..   & 18.75  & 0.14  &   0.52  & 0.31   \\
13 &   0.11 &   1.88 &    ..   &  ..   &    ..   &  ..   & 19.03  & 0.21  &    ..   &  ..    \\
14 & --0.90 & --1.71 &    ..   &  ..   &    ..   &  ..   & 19.04  & 0.22  &    ..   &  ..    \\
15 &   0.65 &   1.37 &    ..   &  ..   &    ..   &  ..   & 19.42  & 0.27  &   0.11  & 0.53   \\ \hline
\multicolumn{11}{l}{$^a$ The positions are relative to star 2 as
determined from the March 18, 1998 observations.} \\
\multicolumn{11}{l}{Star 2 is located at RA = $16^h34^m01\fs 584 \pm 0\fs
020$, Dec = $-47\degr23\arcmin 34\farcs 80 \pm 0\farcs 20$ (J2000)} \\
\multicolumn{11}{l}{$^b$ Magnitudes relative to the March 18, 1998
observations.} \\
\multicolumn{11}{l}{$^c$ To compare these magnitudes with other
observations, the error of 0.054 mag in the} \\
\multicolumn{11}{l}{photometric zeropoint should be added quadratically (see
text)}
\end{tabular}
\end{center}
\end{table*}

A region of $\sim$45\arcsec$\times$45\arcsec , approximately centered on the
radio error circle, was extracted from the resulting frame for each night
and analyzed with DAOPHOT (Stetson \cite{stetson87}) using a constant
PSF. Photometric calibration was obtained through the observation of the
NICMOS standard 9172 (= S279$-$F; Persson et al. \cite{nicmos98}) which was
observed close in time and in position on the sky to the March 14, 1998
observations. The photometric zeropoint of the data was determined by
performing aperture photometry of the (relatively isolated) stars used to
determine the PSF after having removed nearby stars using standard routines
supplied in DAOPHOT. The resulting zeropoints showed a rms spread of 0.054
mag when averaged over the different PSF stars, and we take this value as
the error in the zeropoint. The data in the other nights were calibrated by
comparing the magnitudes of the PSF stars as derived by DAOPHOT in the
different night with the data of March 14, 1998. The resulting {\em
relative} zeropoints show an rms spread of 0.022 mag for the March 12, 1998
data, 0.011 mag for the March 18, 1998 data, and 0.022 mag for the May 16,
1998 data. These values determine the accuracy with which we can compare the
data from the different nights.

\subsection{Variability}

\begin{figure}[h]
\resizebox{\hsize}{!}{\includegraphics{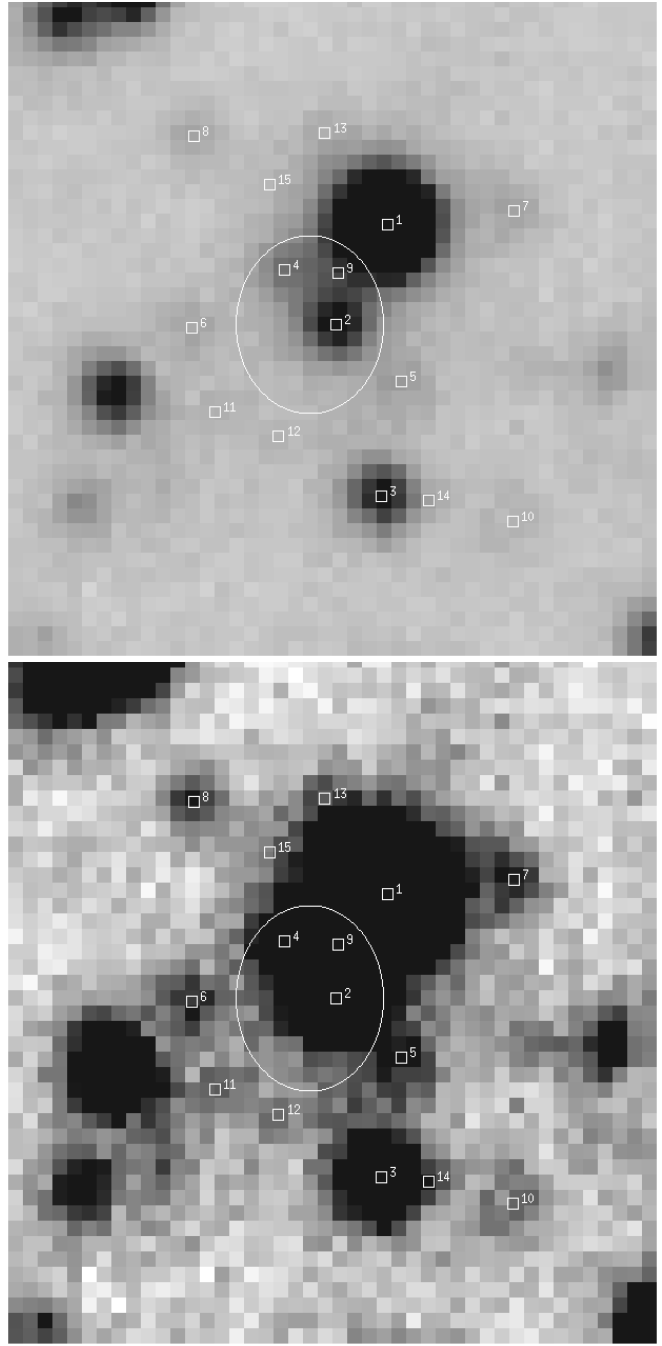}}
\caption[]{ This is an enlarged version of an area
$\sim$6\arcsec$\times$6\arcsec\ around the ATCA error circle presented in
Fig.~\ref{fig:irvar} of the March 18, 1998 $K$ band images. Indicated are
the 15 objects detected by DAOPHOT in or close to this circle (see
text). The image is shown twice with different scaling to show the brighter
sources (top) and fainter sources (bottom)}
\label{fig:sources}
\end{figure}

The main problem in comparing the data from different nights is that due to
the varying seeing, objects which are identified as single objects can split
in to various components, or individual sources can no longer be separated
and are detected as one object. This is especially true for the field of
4U\,1630$-$47 as it is located in the Galactic plane, and the field is very
crowded. As the best seeing was obtained on March 18, 1998, we have used
those data as a basis. In the region of the radio position there is a fairly
tight cluster of stars, and, using DAOPHOT, we identify a total of 15
objects in, or close to, the radio error circle as shown in
Fig.~\ref{fig:sources}. The relative positions and magnitudes of these 15
stars are given in Table~\ref{tab:IRphot}.

Using the position of star 1, which is the brightest of all the stars (see
Table~\ref{tab:IRphot}), as reference point the positions of the 15 stars
were used to determine the magnitudes of the objects in the data from the
other nights. It still could be that an object detected in the March 18,
1998 image in fact consists of more than one star, but in this way we assure
that each time the brightness of the same star(s) is measured in the
different nights. As the seeing during the other nights was significantly
worse, not all 15 stars are detected in each observation. As one might
expect mostly the fainter stars were not detected. To check our method we
also derived the position and magnitude of the sources in the different
nights leaving the positions free. In all cases half of all the sources were
recovered, mostly the brighter objects. The remaining fits converged to
nearby, relatively bright stars. The position and magnitudes of the objects
that were recovered were all consistent with those derived using the fixed
positions, giving us confidence that our method provides reliable results.

In Table~\ref{tab:IRphot} we give the relative magnitudes with respect to
the March 18, 1998 observations of the sources that were detected in each
night using fixed positions. The errors in the relative magnitudes refer to
the formal errors derived by DAOPHOT added quadratically to the error in the
relative zeropoints for each night (see Sect.~\ref{subsect:phot}). Looking
at the last 3 nights, one can see that only star 2 is variable by more than
3-sigma, where the source is 0.215 mag fainter (6.7-sigma) on March 14, 1998
and 0.198 mag fainter (4.8-sigma) on May 16, 1998 with respect to the March
18, 1998 observations. In the March 12, 1998 observations there are various
sources which have varied between 3- and 4-sigma with respect to the March
18, 1998 observations, but we believe this is due to the poor seeing of
these observations which leads to some systematic variation in the measured
brightness of the sources. However, we can conclude from the March 12, 1998
observations that star 2 is at a similar brightness level as during the
March 18, 1998 observations.

There might be some concern about these results given the crowded nature of
the field, and considering the slightly deviating results in the March 12,
1998 observations. To quantify the degree to which seeing contributes to
systematic variations in the measurements we have degraded the data of March
18, 1998 to various degrees by convolving the original data with a 2-d
Gaussian. This was done in such a way as to match the seeing of the other
data sets. Subsequently, the convolved data were then subjected to the whole
reduction process in the same way as the original data. The resulting
magnitudes for the different stars in the experimental data follow fairly
closely the results presented in Table~\ref{tab:IRphot}, with the exception
of star 2 which shows variations far smaller than the variations observed in
the real data. {The results for the simulated data are consistent with
star 2 being constant, but in principle they are also consistent with a
possible systematic shift in the results relative to the March 18, 1998
observations of up to $\sim$0.04 mag. If such systematic shifts exist in the
analysis of the real data, the significance of the observed variations is
somewhat lower}. This analysis does confirm our belief that the relatively
large deviations observed in the results from March 12, 1998 are due to the
poor seeing of the data, and that the observed variations of star 2 are not
just the result of variations in the seeing.

Furthermore, to get a perhaps clearer, or at least a different, measure of
the variability in all images, we formed difference images with the
best-seeing, March 18, 1998 image. For this purpose, we used the optimal
image subtraction technique introduced by Alard \& Lupton (\cite{alard98}).
In this method, the image with better seeing is convolved with a kernel
chosen such that the convolved point-spread function is as close as possible
to that of the image with the worse seeing. The difference images found
using the above method are shown in the lower panels of Fig.~\ref{fig:irvar}
(here, the difference is in the sense image minus convolved March 18, 1998
image). Although the subtraction is not perfect (especially for the
brightest objects) there clearly is a residue close to the center of the
ATCA error circle in the March 14, and May 16, 1998 images (see
Fig.~\ref{fig:irvar}). The centroid of these residues coincide with the
position of star 2, and the relative brightness variations in the different
images are consistent with the photometric brightness variations derived
above, which indicates that these results are not an artifact of the method
used to analyze the data.

Given its variability and position, we conclude that star 2 is the IR
counterpart to 4U\,1630$-$47. From our astrometry we derive a position of RA
= $16^h34^m01\fs 584 \pm 0\fs 020$, Dec = $-47\degr23\arcmin 34\farcs 80 \pm
0\farcs 20$ (J2000) for this star.

\section{Discussion}

\subsection{Previous observations}

Previous IR imaging of the region 4U\,1630$-$47 taken during the 1996
outburst, and in quiescence in 1995 and 1998 were presented by CMG. These
observations were taken with a pixel scale of 0\farcs 65 per pixel and with
a relatively poor seeing of $\sim$1\farcs 5 which makes the detection and
measurement of fainter stars in this crowded field very difficult. In fact,
CMG detect only 2 stars close to the ATCA error circle where we detected 15
sources in our best seeing image. These two stars correspond to stars 1 and
3 as shown in Fig.~\ref{fig:sources} and presented in
Table~\ref{tab:IRphot}. The magnitude of $K$=13.65$\pm$0.05 derived by CMG
for star 1 is fully consistent with our results, while the two measurements
of $K$=16.5$\pm$0.2 and 16.9$\pm$0.2 presented by CMG for star 3 are
somewhat fainter than our results (see Table~\ref{tab:IRphot}). A third star
to the East of star 3 was also measured twice by CMG at $K$=16.8$\pm$0.2 and
16.7$\pm$0.2 mag. For this star we derived a magnitude of 16.02$\pm$0.06,
which is nearly the same as we measured for star 3 and brighter by a similar
amount compared to CMG. As it seems unlikely that these two stars have
changed in brightness by a similar amount, we believe that the measurements
for these stars presented by CMG were affected by the poor resolution (both
in seeing and pixel size) of their data, and that these sources are probably
constant.

Of course, any measurement of star 1 will have included star 2 which we
believe is the counterpart of 4U\,1630$-$47. However, if this star would
have been as bright as during our observations and have contributed all its
flux to the measurement of star 1, it would only contribute 0.1 mag. As CMG
used PSF fitting to derive their magnitudes, star 2 would only have effected
the flank of star 1 on one side, and its contribution will likely have been
a few 0.01 mag at most. It is, therefore, perfectly plausible that star 2
reached a similar brightness in $K$ during the 1996 outburst as during the
1998 outburst.

\subsection{What kind of system is 4U\,1630$-$47?}

Although it is clear that the interstellar reddening towards 4U\,1630$-$47
is high, the precise value is not certain. $N_{\rm H}$ column densities as
derived from spectral fits to X-ray data are in the range
5-15$\times$10$^{22}$\,cm$^{-2}$ (Parmar et al. \cite{parmar86},
\cite{parmar97}; Kuulkers et al. \cite{gnotor98a}; Tomsick et al.
\cite{tomsick98}; Oosterbroek et al. \cite{tim98}; Cui et al. \cite{cui00};
Dieters et al. \cite{stefan00}; Tomsick \& Kaaret \cite{tomsick00};
Trudolyubov et al. \cite{trudo01}) which varies with the X-ray brightness of
the system, and clearly includes a contribution from within the system.
Deriving the column density from HI radio surveys (Kerr et al.
\cite{kerr86}) yields a value of $\sim$2$\times$10$^{22}$\,cm$^{-2}$.
However, at high optical depth the derived column densities are unreliable,
and probably underestimate the true value (Dickey \& Lockman
\cite{dickey90}). Furthermore, this value does not include the possible
contribution of molecular hydrogen. Considering this, we will assume a
conservative range of 2$\times$10$^{22}$\,cm$^{-2}$ $<\,N_{\rm H}\,<$
5$\times$10$^{22}$\,cm$^{-2}$ for the column density towards 4U\,1630$-$47.
Using the relation between $N_{\rm H}$ and interstellar extinction $A_{\rm
V}$ (Predehl \& Schmitt \cite{nhav95}), and a ratio $A_{\rm K}/A_{\rm V} =
0.112$ (Rieke \& Lebofsky \cite{akav85}) the observed column density
corresponds to a $K$ band extinction of 1.3 $<\,A_{\rm K}\,<$ 3.1 mag.

Given its location on the sky and the high interstellar absorption, the
source is most likely at the distance of the Galactic center or beyond, and
following previous authors we will assume a distance of 10 kpc. It is
interesting to note that the source lies in the direction of a giant
molecular cloud (GMC: Corbel et al. \cite{corbel99}). This GMC is located at
a distance of 11 kpc, and the column density of this cloud alone exceeds the
value derived from X-rays, indicating that the source is most likely on the
near side of this cloud. In any case, given the large uncertainty in the
extinction, the precise distance to the source is not very important for the
following discussion. For an average magnitude of $K$ = 16.1 for star 2
during our observations we derive --0.2 $>\,M_{\rm K}\,>$ --2.0 mag.

The general X-ray properties of 4U\,1630$-$47 suggest that the source is a
(transient) low-mass X-ray binary (LMXB), and the optical/IR emission in
outburst is likely dominated by the hot X-ray heated accretion disk (i.e.,
$T>10^4$ K over most of the disk surface; see Van Paradijs \& McClintock
\cite{jvpmcc95}). For \object{Sco X-1}, which is thought to be dominated by
emission from its accretion disk $(V-K)_0=-0.3$ (Hertz \& Grindlay
\cite{hertz84}; Van Paradijs \& McClintock \cite{jvpmcc94}, henceforth
VPM). However, in the case a system contains a late-type giant secondary it
is expected to contribute significantly to the IR emission compared to the
optical, and for \object{GS\,2023+338} in outburst we derive $(V-K)_0=0.6$
(Gehrz et al. \cite{gehrz89}; VPM). Using these two values as extremes, we
derive for the visual absolute magnitude of 4U\,1630$-$47 in outburst 0.4
$>\,M_{\rm V}\,>$ --2.3 mag.

The absolute visual magnitudes of LMXBs with known distances and orbital
periods are well represented by the relation $M_{\mathrm{V}} = 1.57 (\pm
0.24) -2.27 (\pm 0.32)~\log~\Sigma$, where $\Sigma =
\gamma^{1/2}~(P_{\mathrm{orb}}({\mathrm{hr}}))^{2/3}$, and $\gamma$ is the
observed X-ray luminosity in units of the Eddington limit for a 1.4
M$_{\odot}$ neutron star (VPM). The average X-ray flux at the time of the IR
observations was $2.8\times 10^{-9}$\,erg\,cm$^{-2}$\,s$^{-1}$ (see
Fig.~\ref{fig:xrlc}), which corresponds to $8.5\times
10^{36}$\,erg\,s$^{-1}$ at 10 kpc. The above relation then implies 11 hr
$\la P_{\mathrm{orb}} \la$ 29 d. Assuming that the outbursts in SXTs are due
to a disk instability and taking into account the effect of X-ray heating of
the disk Van Paradijs (\cite{jvp96}) derived that for a system to show
outbursts the X-ray luminosity must conform to the inequality $\log
\overline{L_{\mathrm{X}}} < 35.8 + 1.07~\log~P_{\mathrm{orb}}(\mathrm{hr})$.
The average X-ray luminosity over the outburst cycles for the 1996 and 1998
outbursts (CMG; Oosterbroek et al. \cite{tim98}) corresponds to
$\overline{L_{\mathrm{X}}} = 1\times 10^{37}$\,erg\,s$^{-1}$ (at 10 kpc),
and for 4U\,1630$-$47 to be transient its orbital period must be
$P_{\mathrm{orb}} \ga$ 13 hr which agrees well with the above estimate.

Compared to other black-hole SXTs, the orbital period estimates best match
those of 4U\,1543$-$47, \object{XTE\,J1550$-$564}, GRO\,J1655$-$40,
SAX\,J1819.3$-$2525 and GS\,2023+338 (Orosz et al. \cite{orosz98},
\cite{orosz01}; S\'anchez-Fern\'andez et al. \cite{sanchez99}; Jain et
al. \cite{jain01b}; Shahbaz \& Kuulkers \cite{shahbaz98}, and references
therein). Also the estimated absolute visual magnitude for 4U\,1630$-$47
agrees well with that of 4U\,1543$-$47 and SAX J1819.3$-$2525, though it is
too bright compared to XTE\,J1550$-$564, and a bit faint compared to
GRO\,J1655$-$40 and especially GS\,2023+338.

The main problem with the identification of the source as an LMXB is the
small observed brightness variations. For an X-ray heated disk one expects
the flux to vary as $F \propto F^{0.5}_{\mathrm{X}}$ (Van Paradijs \&
McClintock \cite{jvpmcc95}). During our IR observations the X-ray flux
varied by a factor $\sim$5 (see Fig.~\ref{fig:xrlc}), which implies a
variation of $\sim$1 mag in the brightness of the disk as compared to the
small variation observed over the course of our observations. However, if
the source is a long orbital period LMXB, the secondary is expected to
contribute significantly to the brightness in the IR reducing the expected
variation seen in the $K$ band. Furthermore, the outburst amplitude in the
optical (and most likely the IR as well) of 4U\,1543$-$47 and
GRO\,J1655$-$40 are the smallest among SXTs (Orosz et al. \cite{orosz98};
Shahbaz \& Kuulkers \cite{shahbaz98}) and they varied only by $\sim$2-3 mag
over the course of an entire outburst. By analogy one might expect that any
variation in average brightness in the IR of 4U\,1630$-$47 during an
outburst is small.

The latter two sources (and also SAX\,J1819.3$-$2525) all contain rather
early-type (late-B to mid-F) secondaries, and to a certain extent they are
not true `low-mass' X-ray binaries. However, the mass-transfer in these
source is still thought to occur through Roche-lobe overflow and the optical
emission (at least in outburst) is dominated by the (X-ray heated) accretion
disk. In that sense, the above discussion concerning the orbital period of
4U\,1630$-$47 is still valid.

The detection of a dip in the X-ray light curve during the 1996 outburst
(Kuulkers et al. \cite{gnotor98a}) indicates that the inclination of the
system is fairly high ($i\sim 60-75\deg$), similar to GRO\,J1655$-$40. In
the latter system the orbital brightness variations have an amplitude of
$\sim$0.3--0.5 mag (Van der Hooft et al. \cite{vdh97}, \cite{vdh98}), which
could also mask longer term average brightness variations if the source is
observed on only a few occasions as is the case for our observations of
4U\,1630$-$47. Another possibility is that there is an additional (varying)
source of IR emission. In fact, the May 16, 1998 observations coincide with
the soft-to-hard state transition observed in X-rays (Tomsick \& Kaaret
\cite{tomsick00}). In other systems in the low/hard X-ray state there is an
excess radio emission component which probably extends to the IR (Fender et
al. \cite{fenderetal01}; Fender \cite{fender01} and references therein), and
most likely originates from a compact jet. Especially interesting in that
respect are the observations of XTE\,J1550$-$564 during outburst (Jain et
al. \cite{jain01a}). The transition from the soft to the hard state itself
has also been associated with the ejection of relativistic plasma (Corbel et
al. \cite{corbel00}; Corbel et al. \cite{corbel01}; Brocksopp et
al. \cite{brock01}).

We conclude that our observations are consistent with 4U\,1630$-$47 being a
system with a relatively long orbital period, and that it is most likely
similar to 4U\,1543$-$47, GRO\,J1655$-$40 or SAX\,J1819.3$-$2525 containing
a relatively early type secondary.

Not only do the general properties of 4U\,1630$-$47 suggest that it is a
LMXB type system, also GRO\,J1655$-$40 and \object{GRS\,1915+105}, which are
the two sources that most closely resemble its X-ray behavior (e.g.,
Kuulkers et al. \cite{gnotor97a}, \cite{gnotor97b}), are both LMXBs (Bailyn
et al. \cite{bailyn95}; Greiner \cite{greiner01}). However, we note that the
range in absolute $K$ band magnitude derived for 4U\,1630$-$47 also agrees
with the source being a (intrinsically slightly reddened) B star (at 10
kpc), which one would expect if the source is a detached Be/X-ray binary.
Furthermore, the observed variability is typical for what is observed for
these binaries in the IR (Coe et al. \cite{coe97}), so on the basis of our
photometry alone we can not exclude this possibility. Probably only IR
spectroscopy will be able to settle this question definitely.

\begin{acknowledgements}
We thank Chris Lidman for help with obtaining the observations. EK thanks
John Tomsick for providing their XSPEC spectral fit results in numerical
form. The RXTE/ASM data used in this paper was obtained from the public ASM
database at MIT, which is supported by the ASM and RXTE instrument teams at
MIT and Goddard Space Flight Center. We thank the referee John Tomsick for
various useful comments.
\end{acknowledgements}

\end{document}